\documentclass[final,twocolumn,aps,floatfix]{revtex4}

\usepackage{t1enc}
\usepackage[latin1]{inputenc}
\usepackage{amsmath}
\usepackage{amssymb}
\usepackage{showkeys}
\usepackage{graphicx}
\usepackage{citesort}
\usepackage[dvips]{color}

\newcommand{\cE}{{\mathcal E}}
\newcommand{\cH}{{\mathcal H}}

\newcommand{\cL}{{\mathcal L}}

\newcommand{\cO}{{\mathcal O}}

\newcommand{\CC}{{\mathbb C}}

\newcommand{\ZZ}{{\mathbb Z}}

\newcommand{\2}{\tfrac12}
\newcommand{\del}{\partial}
\renewcommand{\i}{\mathrm{i}}
\newcounter{mnotecount}[section] %

\begin{document}

\title{On discretizations of axisymmetric systems}

\author{J\"org Frauendiener}
\affiliation{Institut f\"ur Theoretische Astrophysik,
Universit\"at T\"ubingen,
Auf der Morgenstelle 10,
D-72076 T\"ubingen,
Germany}
\email{joergf@tat.physik.uni-tuebingen.de}
\homepage{http://www.tat.physik.uni-tuebingen.de/~joergf}

\begin{abstract}
In this paper we discuss stability properties of various
discretizations for axisymmetric systems including the so called
cartoon method which was proposed by Alcubierre, Brandt et.al. for the
simulation of such systems on Cartesian grids. We show that within the
context of the method of lines such discretizations tend to be
unstable unless one takes care in the way individual singular terms
are treated. Examples are given for the linear axisymmetric wave
equation in flat space.
\end{abstract}
\maketitle

\section{Introduction}
\label{sec:intro}

Axisymmetric systems are notorious for the difficulties they pose in
numerical simulations when they are expressed in coordinates adapted
to the symmetry. The problem is due to the singular nature of the
coordinates on the axis, i.e., the set of fixed points for the
symmetry transformations or, equivalently, the set where the
infinitesimal generator of the symmetry -- the Killing vector --
vanishes. Adapted coordinates, i.e., an angle $\phi$ along the Killing
vector, a radius $r$ measuring `distance' from the axis and a third
coordinate $z$, yield coordinates $(r,z)$ on the space of orbits of
the action of the symmetry group. The general theory of group
actions~\cite{jaenich68:_differ_g_mannig,frauendiener87:_es_raumz_flues}
yields the following facts. The orbit space contains two types of
orbits\footnote{We ignore here the so called `exceptional orbits'
because we are only interested in the neighbourhood of the axis.}: the
`regular orbit' is a circle around the axis. It has maximal dimension
(one) and the union of all regular orbits is dense in the space of
orbits. The other type of orbit is a fixed point which has dimension
zero. The set of fixed points is a submanifold of codimension two, the
`axis'.  The drop in dimension of the orbits shows up in the topology
of the space of orbits which acquires a boundary corresponding to the
axis points and in a degeneracy of the adapted coordinates: since on
the axis (for fixed $z$) all values of $\phi$ address the same point
the angular coordinate becomes irrelevant on the axis, so that the
coordinate system degenerates there.

Invariant tensor fields transform in a rigid way (Lie transport) under
the symmetry so that they are determined on an entire regular orbit
once they are known on one of its points. Therefore, in order to
determine the fields on the entire space it is enough to determine
them on a hypersurface which is transversal to the orbits. Thus, the
problem is reduced to a problem in a space with lower dimension. For
invariant tensor fields at a fixed point the invariance implies a
transformation law among the tensor components which must be
interpreted as a regularity condition for the tensor field on the
axis.

The advantage of using the adapted coordinates is, of course, the
manifestation of this reduction. In these coordinates the components
of the tensor fields do not depend on the angular coordinate along the
Killing vector. Thus, there are only two essential coordinates left
over and the problem reduces in complexity because an axisymmetric
problem on some open 3-dimensional domain reduces to an equivalent
2D-problem on a domain with boundary. The boundary conditions are
obtained from regularity conditions on the axis.

The disadvantage of using the adapted coordinates is that the reduced
equations become formally singular on the axis. The regularity
conditions on the axis guarantee that the individual terms in the
equation have in fact a regular limit on the axis so there is not a
real problem. Yet, numerically, one has to evaluate terms which are
formally $0/0$ or $\infty - \infty$ which does pose problems. Thus, it
seems that one has the choice between a 3D system with regular, such
as Cartesian, coordinates and regular equations or a 2D system with
degenerate coordinates and singular equations.

In an attempt to combine the best of both worlds, Alcubierre
et~al.~\cite{alcubierrebrandt99:_symmet} proposed what they called the
cartoon method. This scheme was designed to borrow from the
singularity-free nature of full 3D Cartesian coordinates $x$, $y$, $z$
and to allow the treatment of axisymmetric problems without the memory
constraints of a full 3D problem. The method uses the Killing
transport equation along the orbits to find the field values on the
grid points of a Cartesian grid from its values on the hypersurface
$y=0$. These values are obtained by interpolation from the grid points
lying in that hypersurface. For a second order method one needs
to keep only three hypersurfaces $y=const$ in memory so that the
problem scales quadratically with the number of points and not
cubically. 

The plan of the paper is as follows. In section~\ref{sec:method} we
discuss the cartoon method in more detail. In order to have a specific
problem at hand we apply it to the axisymmetric linear wave equation
in section~\ref{sec:axiwave}. We write this equation as a first order
symmetric hyperbolic system. The reason for this apparent complication
is that we have implemented this method for the axisymmetric conformal
field equations which are a symmetric hyperbolic system and because
the linear wave equation served as a model equation to test various
implementations~\cite{frauendienerhein02:_numer}. It turns out that
the crucial ingredient to the method is the interpolation and we
discuss various possibilities. Furthermore, we show that there is no
real difference between the cartoon method using Cartesian coordinates
and the formulation in adapted coordinates so that the issues which
are relevant for the former apply to the latter as well. In
section~\ref{sec:numerics} we analyse the stability properties of the
cartoon method with different interpolation procedures as well as
other possible discretizations when the method of lines is used for
time evolution. We end with a brief discussion of the results.

\section{The Cartoon method}
\label{sec:method}

In this section we present the cartoon method in more detail. We
consider a 3-dimensional Euclidean space $\cE^3$ with Cartesian
coordinates $\mathbf{x}=(x,y,z)$ on which we define the action of the circle
group $SO(2) = G$
\[
G = \left\{ g =\left(\begin{smallmatrix} a & -b\\ b&a
    \end{smallmatrix}\right) \mid a^2 + b^2 =1\right\}
\] 
in the usual way by
\begin{equation}
  \label{eq:action}
  \begin{split}
    \Phi: \left(G,\cE^3\right) &\to \cE^3,\\
    \left( g, \mathbf{x}
    \right) &\mapsto \Phi(g,\mathbf{x}) = (a x - by, bx + ay,z) =
    g\cdot\mathbf{x}.
\end{split}
\end{equation}
The infinitesimal generator of this action is the vector field $\xi =
x\del_y - y\del_x$ which vanishes on the axis $x=y=0$, the set of
fixed points for the action. The action $\Phi$ is the prototype for
any axisymmetric system in the neighbourhood of the axis in the sense
that for any of these systems one can find local coordinates in which
the action takes the above
form~\cite{jaenich68:_differ_g_mannig,frauendiener87:_es_raumz_flues}.
Therefore, we will focus here on flat space because we are mainly
interested in the local properties in a neighbourhood of the axis.
Curvature will not play an essential role.

Introducing adapted cylindrical coordinates $(r,\phi,z)$ the
coordinate expression for the action is
\[
\left( \left(
  \begin{smallmatrix}
    \cos \theta& -\sin \theta\\
    \sin \theta&  \cos \theta
  \end{smallmatrix} \right), 
(r,\phi,z)\right) \mapsto (r, \phi+\theta,z)
\]
and the Killing vector is given by
\[
\xi = \del_\phi.
\]
These coordinates are singular on the axis which can be seen e.g.,
from the fact that the Jacobi matrix of the transformation between
cylindrical and Cartesian coordinates has vanishing determinant there.

A tensor field $T$ in $\cE^3$ is axisymmetric if for any $g \in G$ it
satisfies the condition
\begin{equation}
\Phi_g^* T = T\label{eq:invPhi}
\end{equation}
where $\Phi_g^*$ denotes the pull-back with $\Phi_g: \cE^3 \to \cE^3,
\mathbf{x} \mapsto g \cdot \mathbf{x}$. This condition expresses the
invariance of $T$ with respect to the action of $SO(2)$. The
infinitesimal version of this condition is the equation of Killing
transport
\begin{equation}
\cL_\xi T = 0.\label{eq:invKill}
\end{equation}
Suppose we want to solve a quasi-linear system of PDEs of the form
\begin{equation}
\dot T = F(\del_a T,T,t,\mathbf{x})\label{eq:pde}
\end{equation}
where $T$ is some given time dependent axisymmetric tensor field
on~$\cE^3$. Here, $\del_a$ denotes covariant differentiation.  The
usual way to treat this system is to express it in terms of the
adapted coordinates and to use the invariance
condition~\eqref{eq:invKill} which asserts, that the components of $T$
in the basis of adapted coordinates do not depend on $\phi$. Thus, the
essential coordinates are $r$ and $z$ and the problem is reduced to a
2-dimensional one. However, from the form of the covariant derivative
operator, i.e., the Christoffel symbols of the metric expressed in
cylindrical coordinates it is clear, that the equation will, in
general, contain singular terms.

A different procedure is to employ any coordinate system which is
regular in a neighbourhood of the axis and then to make explicit use
of the invariance condition~\eqref{eq:invPhi} or its infinitesimal
version~\eqref{eq:invKill}~\cite{thornburgh:pc}. So we may choose
Cartesian coordinates with the axis at $x=y=0$. Let $T$ be an
arbitrary axisymmetric tensor field on $\cE^3$. From the
invariance condition~\eqref{eq:invPhi} we can find its values at any
point in $\cE^3$ from its values on the half-plane $\cH =
\{y=0,x\ge0\}$. To illustrate this let us first assume that $T$ is a
scalar field. From the invariance condition we have
\[
T(\mathbf{x}) = T(g\cdot\mathbf{x})
\]
for any rotation matrix $g$ and each point $\mathbf{x}=(x,y,z)$. Thus, in
particular, choosing 
\[
g = 
\begin{pmatrix}
  \frac{x}{\sqrt{x^2+y^2}} &  \frac{y}{\sqrt{x^2+y^2}} \\
  \frac{-y}{\sqrt{x^2+y^2}} &  \frac{x}{\sqrt{x^2+y^2}} 
\end{pmatrix}
\]
for $x\ne0$, $y\ne0$ we have $\Phi_g(\mathbf{x}) = (r,0,z)$ with
$r=\sqrt{x^2+y^2}$ so that
\begin{equation}
T(\mathbf{x}) = T(r,0,z).\label{eq:scalar}
\end{equation}
Similarly, if $T=T_1\,dx + T_2\,dy + T_3\,dz$ is a 1-form, we have
\[
\Phi_g^* T(\mathbf{x}) = T_1(g\cdot \mathbf{x})\,d(\Phi_g^*x) + T_2(g\cdot
\mathbf{x})\,d(\Phi_g^*y) + T_3(g\cdot \mathbf{x})\,d(\Phi_g^*z).
\]
Inserting the explicit form of the matrix $g$ yields
\begin{widetext}
\begin{multline*}
  \Phi_g^* T(\mathbf{x}) = T_1(g\cdot \mathbf{x})\,\left( a\,dx -
    b\,dy\right) + T_2(g\cdot \mathbf{x})\,\left(b\,dx + a\,dy\right)
  + T_3(g\cdot \mathbf{x})\,dz\\
  =\bigl(a T_1(g\cdot \mathbf{x}) + b T_2(g\cdot \mathbf{x})\bigr)\,dx
  +\bigl(a T_2(g\cdot \mathbf{x}) - b T_1(g\cdot \mathbf{x})\bigr)\,dy
  + T_3(g\cdot \mathbf{x})\,dz
\end{multline*}
\end{widetext}
so that, in this case, the invariance condition implies the three equations
\begin{gather*}
  T_1(\mathbf{x}) = a T_1(g\cdot \mathbf{x}) + b T_2(g\cdot
  \mathbf{x}),\\ 
  T_2(\mathbf{x}) = -b T_1(g\cdot \mathbf{x}) + a
  T_2(g\cdot \mathbf{x}),\\ 
  T_3(\mathbf{x}) = T_3(g\cdot\mathbf{x}).
\end{gather*}
Choosing, as before, $a=x/\sqrt{x^2+y^2}$ and $b=-y/\sqrt{x^2+y^2}$,
we get the relations
\begin{gather}
  T_1(\mathbf{x}) = \frac{1}{r}\,
  \left(x\, T_1(r,0,z) - y\, T_2(r,0,z)\right),\nonumber\\ 
  T_2(\mathbf{x}) = \frac{1}{r}\, 
  \left(y\,T_1(r,0,z) + x\, T_2(r,0,z)\right),\label{eq:oneform}\\ 
  T_3(\mathbf{x}) = T_3(r,0,z).\nonumber
\end{gather}
Similar relations hold for tensor fields with different type. This
shows that we can get information about the values at all points
outside the axis from values at the points on $\cH$ except for the
axis. \emph{On} the axis these relations imply regularity conditions
for the tensor fields. E.g., in the case of a 1-form above we get
$T_1(0,0,z)=T_2(0,0,z)=0$ which follows by taking limits from
various directions towards the axis points.

How are we to use this analytical background in numerical
applications? Following~\cite{alcubierrebrandt99:_symmet} we cover the
half-plane $\cH$ by a regular Cartesian grid
\[
x_i=i\Delta x, z_k=k\Delta z, \text{ with } i,k \in \ZZ, i\ge0
\]
which we think of as being a part of a full 3D cartesian grid with
points given by $x_i,y_j,z_k$ with $y_0=0$.  Each component of a
tensor field $T$ yields a grid function $T_{ijk}$. The axisymmetry
implies that we should be able to determine the functions $T_{ijk}$
from their values on $\cH$ alone.

In order to treat equations like~\eqref{eq:pde} above one needs to
compute approximations to the spatial derivatives of the
components. This is straightforward in the case of $x$- and
$z$-derivatives. However, for the $y$-derivatives we need to find the
values of the grid functions at points outside~$\cH$.
\begin{figure}
  \begin{center}
    \includegraphics[width=0.4\textwidth]{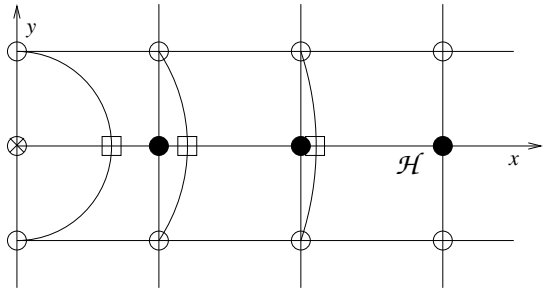}
    \caption{\label{fig:cartoon} Illustration of the method (see text)}
  \end{center}
\end{figure}
The situation is illustrated in Fig.~\ref{fig:cartoon}. It shows the
view from the positive $z$-axis down onto the $(x,y)$-plane. The black
dots indicate grid points in $\cH$, the point on the left with the
cross being the axis. The open circles indicate points outside $\cH$
with $y=\pm \Delta y$. The values of the fields at these points have
to be computed from the invariance
equation~\eqref{eq:invPhi}. Following the orbits through these outside
points one can connect them with points in $\cH$, indicated by the
open squares, and one can find the values at the outside points from
the values at these points on $\cH$. Thus, e.g.,
applying~\eqref{eq:scalar} in this situation we obtain for a scalar
\[
T(x_i, \pm\Delta y,z_k) = T(\sqrt{x_i^2 + \Delta y^2},0,z_k),
\]
while the corresponding transformation for a 1-form follows
from~\eqref{eq:oneform}. Note, that the squares are on $\cH$ but they
are not, in general, grid points. Thus, having reduced the problem to
$\cH$ the remaining question is how to determine the values at these
points. The natural choice is to use some form of interpolation as it
is also suggested in~\cite{alcubierrebrandt99:_symmet}. However, it is
not clear what kind of interpolation should be used and it is the
purpose of this work to point out that not all interpolation schemes
perform well.

\section{The axisymmetric wave equation}
\label{sec:axiwave}

To have a concrete example at hand we will focus now on the
axisymmetric scalar wave equation in flat space. It is clear from the
considerations above that the $z$-dependence is irrelevant for this
analysis. Therefore, we simplify even further by assuming that the
field is in fact axisymmetric and $z$-independent. Then the equation
is
\[
\ddot\phi = \phi_{xx} + \phi_{yy}
\]
where $\phi = \phi(t,x,y)$ is axisymmetric. Introducing the 1-form
$d\phi = u\,dt + v\,dx + w\,dy$  we can derive from this equation a
symmetric hyperbolic first order system
\begin{gather}
  \dot u = v_x + w_y,\label{eq:wave1}\\
  \dot v = u_x,\label{eq:wave2}\\
  \dot w = u_y\label{eq:wave3}
\end{gather}
with the constraint $v_y-w_x=0$. The function $\phi$ can be recovered
from a solution of this system by either evolving $\dot \phi = u$ or
by solving $\Delta\phi=v_x+w_y$ at each time step. Since $\phi(t,x,y)
= \phi(t,\sqrt{x^2+y^2},0) = \phi(t,\pm x,\pm y)$ we have, in
particular, $w(t,x,0)=0$ and $v(t,0,0)=0$. Furthermore, taking
derivatives we obtain $u_y(t,x,0)=0$ and $v_y(t,x,0)=0$. Hence, we can
forget about~\eqref{eq:wave3} and the constraint, because these
equations are identically satisfied on~$\cH$. So we are left with the
system
\begin{gather}
  \dot u = v_x + w_y,\\
  \dot v = u_x.
\end{gather}
Before we go into the more numerical details it is worthwhile to write
down the system in adapted (i.e., cylindrical) coordinates $(r,\phi,z)$
\begin{equation}
  \dot u = v_r + \frac{v}{r}, \qquad \dot v = u_r,\label{eq:adapted}
\end{equation}
which follows by expressing the Laplace operator in these coordinates,
using the independence of $z$ and $\phi$ and then introducing the
first derivatives as new dependent variables.

We solve these equations numerically by applying the method of lines,
i.e., we discretize the spatial derivatives to obtain a system of
ODEs. This can then be solved with standard ODE solvers.  We put the
equations on a grid as indicated above. Now $\cH$ is in effect a
1-dimensional grid with points $x_i = i\Delta x$ and we have grid
functions $u_i(t)=u(t,i\Delta x,0)$, $v_i(t) = v(t,i\Delta x,0)$. According
to the idea of the cartoon method we compute the term $w_y$, which of
course does not vanish on $\cH$, by following the orbits. From the
transformation law~\eqref{eq:oneform} for a 1-form we get
\begin{equation}
\label{eq:trans}
  w(t,x,y) = \frac{y}{\sqrt{x^2+y^2}}\;  v(t,\sqrt{x^2+y^2},0).
\end{equation}
The discretization of the term $w_y$ using centered differences to get
a second order accurate approximation yields
\[
w_y(t,x_i,0) = \frac{w(t,x_i,\Delta y) - w(t,x_i,-\Delta y)}{2\Delta y}
= \frac{v_*}{x_*} ,
\]
where $x_*=\sqrt{x_i^2+\Delta y^2}$ and $v_*=v(t,x_*,0)$. This, then,
yields the final system obtained by application of the cartoon method
\begin{equation}
\label{eq:system2nd}  \dot u = v_x + \frac{v_*}{x_*},\qquad  \dot v = u_x.
\end{equation}
It is instructive to compare this system with the
system~\eqref{eq:adapted} obtained in adapted coordinates. Expanding
in powers of $\Delta y$ we get
\[
\frac{v_*}{x_*} = \frac{v(x_*)}{\sqrt{x_i^2 + \Delta y^2}} =
\frac{v_i}{x_i} + \frac12 \left(\frac{\del v}{\del x}(x_i) -\frac{v_i}{x_i} 
  \right) \frac{\Delta y^2}{x_i^2} + \cO(\Delta y^4).
\]
Hence, \emph{up to the accuracy} $\Delta y^2$ for which the derivation
of~\eqref{eq:system2nd} is valid, \emph{the two systems agree}. 

One can, in fact, say even more. The infinitesimal invariance
condition~\eqref{eq:invKill} or, equivalently, taking the $y$-derivative
of~\eqref{eq:trans} and evaluating at $y=0$ yields
\[
w_y(t,x,0) = \frac{v(t,x,0)}{x}
\]
for all $x>0$. Hence, whenever we use the cartoon method to
approximate $w_y$ to a certain order in the grid spacing $\Delta y$ we
approximate $v/x$ to the same order. Thus, there is no difference in
that order of accuracy between the systems~\eqref{eq:system2nd}
and~\eqref{eq:adapted}. And, therefore, the discussions in the
following section apply to both.

\section{Numerical issues}
\label{sec:numerics}

In this section we discuss the stability properties of various
implementations of the systems~\eqref{eq:adapted}
and~\eqref{eq:system2nd}. In the spirit of what was said in the
previous section we regard the term $\frac{v_*}{x_*}$ as an
approximation to $\frac{v}{x}$ which is not necessarily located on a
grid point. Then, the main question is how to approximate this term on
$\cH$. We implement several possibilities and look at their stability
properties.

In order to have a well defined problem we have to worry about
boundary conditions at the origin $x=0$ and at an outer boundary which
we put at $x=1$. The conditions at the origin follow from the
transformation properties of the fields under the symmetry. They are
\begin{gather}
  v(t,-x,0) = -v(t,x,0) \rightarrow v(t,0,0)=0,\\
  u(t,-x,0) = u(t,x,0) \rightarrow \frac{\del u}{\del x}(t,0,0) = 0.
\end{gather}
The fact, that $x=0$ is the fixed point of the symmetry is reflected
in the simultaneous vanishing of $v$ and $x$. This also implies that
at the origin we have $\frac{v}{x}=v_x$ so that at $x=0$ the time
evolution of $u_0=u(t,0,0)$ is given by the equation $\dot u_0 =
2v_x(t,0,0)$.

Since we are interested in the properties of the scheme in the
interior we implement some sort of periodic boundary conditions. Since
we want to avoid a second origin at $x=1$ we imagine that our wave
equation lives on a 2-sphere and is symmetric under rotations of the
sphere around its poles. Now we impose the condition that it be also
symmetric under reflection across the equator. This discrete symmetry
allows us to consider the equation only along a meridian from the
north pole down to the equator. The symmetry implies conditions for
the fields at the equator which we put at $x=1$:
\begin{gather}
  v(t,1-x,0) = - v(t,1+x,0) \rightarrow v(t,1,0) = 0,\\
  u(t,1-x,0) = u(t,1+x,0) \rightarrow \frac{\del u}{\del x}(t,1,0) = 0.
\end{gather}

In the context of the method of lines we write down a system of ODEs
for the grid functions $v_i$ and $u_i$. This will be a linear system
of the form $\dot f = \mathbf{A} f$ and we can find its stability
properties by looking at the spectrum of the matrix $\mathbf{A}$ which
we determine numerically.

With these preparations out of the way we can now write down the first
of our discretizations for the system~\eqref{eq:adapted}. This is the
first one which comes to mind, namely
\begin{equation}
  \begin{aligned}
    \dot u_0 &= \frac{2v_1}{h},\\
    \dot u_i &= \frac{v_{i+1} - v_{i-1}}{2h} + \frac{v_i}{ih},\\
    \dot u_N &= -\frac{v_{N-1}}{h}\\
  \end{aligned}
  \qquad
  \begin{aligned}
    \dot v_0 &= 0,\\
    \dot v_i &=  \frac{u_{i+1} - u_{i-1}}{2h}\\
    \dot v_N &= 0.
  \end{aligned}
\end{equation}
Here we have chosen $h=1/N$ where $N+1$ is the number of grid
points so that $x_i=ih$.  The matrix $\mathbf{A}$ which corresponds to
this discretization has dimensions $2(N+1) \times 2(N+1)$ and the form
\[
\mathbf{A}= \frac1h
\left(
\begin{array}{c|c}
0&\mathbf{C}\\
\hline
\mathbf{D}&0  
\end{array}
\right)
\]
where $\mathbf{C}$ is the matrix
\[
\mathbf{C} = 
\left(
\begin{array}{cccccc}
  0 & 2 & 0 & \cdots&&\\
  -\frac12 & 1 & \frac12 &&&\\
  & -\frac12 & \frac12 & \frac12 &&\\
  &&\ddots&\ddots&\ddots&\\
  &&& -\frac12 & \frac1{N-1} & \frac12\\
   & & \cdots& 0 & -1 & 0
\end{array}
\right)
\]
and $\mathbf{D}$ is 
\[
\mathbf{D} = 
\left(
\begin{array}{ccccc}
  0 & 0  & \cdots&0&0\\
  -\frac12 & 0 & \frac12 &&\\
  &\ddots&\ddots&\ddots&\\
  && -\frac12 & 0 & \frac12\\
  0 & 0  & \cdots&0&0\\
\end{array}
\right).
\]
This system of ODEs is solved by a standard ODE solver, such as a
Runge-Kutta scheme or a multistep scheme like the Adams-Bashforth
schemes. We also consider the `iterated Crank-Nicholson' (ICN) scheme
which has become popular among numerical relativists
(see~\cite{teukolsky99:_crank_nichol_method_numer_relat} for an
analysis of this scheme). Relevant for deciding about the stability of
a difference approximation for ODEs is the region of absolute
stability (see
e.g.~\cite{hairernorsett00:_solvin_ordin_differ_equat,%
hairerwanner96:_solvin_ordin_differ_equat}). This is a closed set
$\Omega$ in the complex plane which consists of those $z=\lambda
\Delta t\in \CC$ for which the method applied to the equation $\dot u
= \lambda u$ with step size $\Delta t$ produces bounded
approximations. The stability regions for the schemes mentioned above
are shown in appendix~\ref{sec:stabreg} and the stability region(s)
for ICN is determined in appendix~\ref{sec:icn}.

Thus, in the context of the method of lines, the stability conditions
can be obtained by determining the eigenvalues of the discretization
matrix $\mathbf{A}$ and checking whether it is possible to scale it so
that it fits entirely into the stability region of the ODE solver. The
scale factor determines the maximal admissible time step. Strictly
speaking, this consideration applies only to the linear case. For
non-linear equations this check should be understood as a `rule of
thumb'~\cite{trefethen00:_spect_matlab}.

\begin{figure}[ht]
  \begin{center}
    \includegraphics[width=7cm]{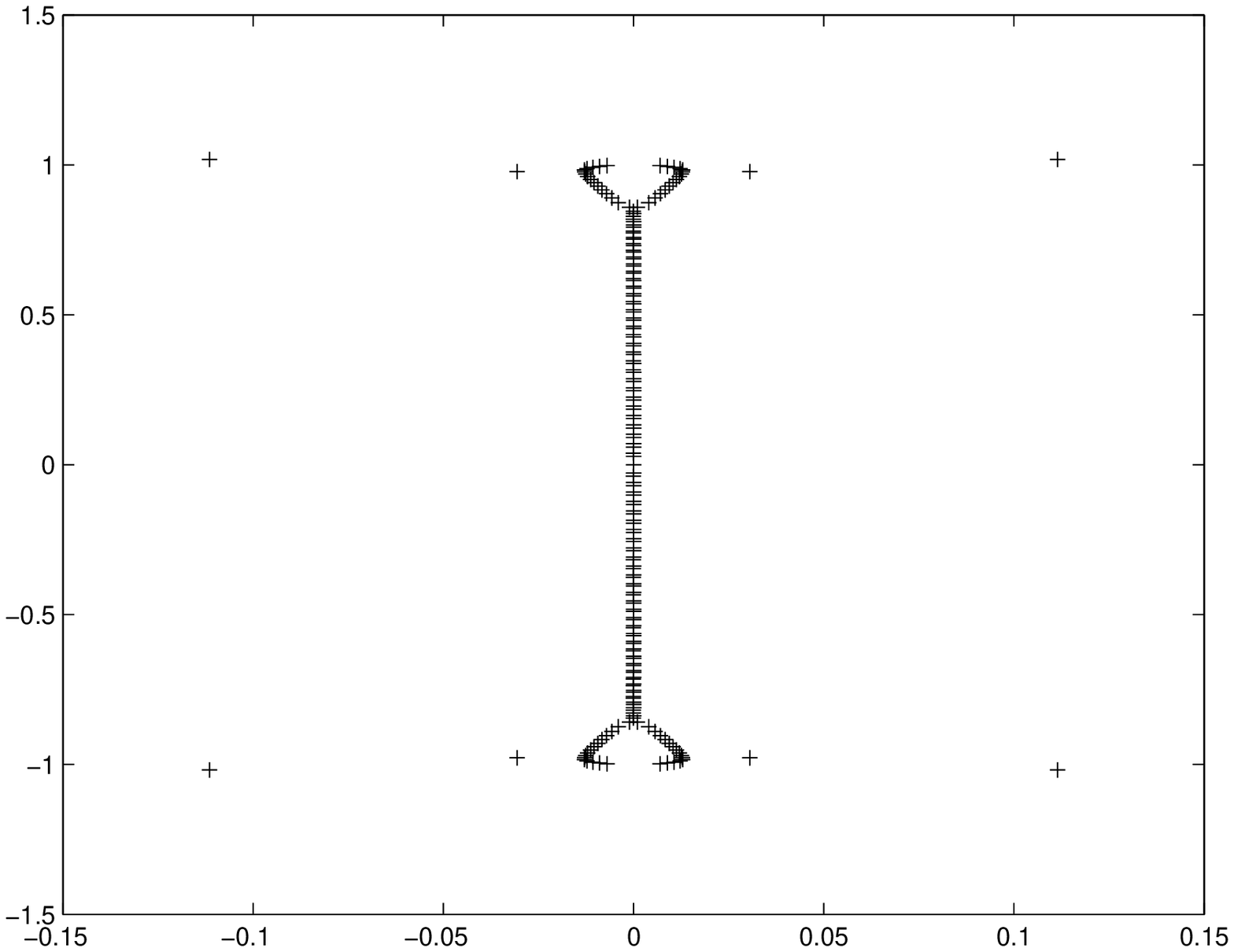}
    \caption{\label{fig:straight}Spectrum of the `straightforward'
      discretisation, scheme 2a}
  \end{center}
\end{figure}

The spectrum $\sigma_N(h\mathbf{A})$ of $h \mathbf{A}$ is
shown for $N=100$ in Fig.~\ref{fig:straight}. 
We point out some of its properties which are valid for all other
discretizations below as well. The spectrum is symmetric with respect
to reflections about the real axis as well as about the imaginary
axis. While the former symmetry is due to the reality of the equation,
the latter is a consequence of the fact that the underlying equation is a
second order wave equation which has in- and outgoing
modes. Furthermore, the spectrum is concentrated mostly on the
imaginary axis with a few `outliers'. It can be seen that the
eigenvalues with the largest, in absolute value, real part remain
unchanged when we change $N$ which implies that the spectral radius of
$\mathbf{A}$ is proportional to $N$.

An interesting feature of the spectrum are the two `handles' which are
located near $\pm \i$. These, and the outliers are due to the $v/x$
term in the equation and, depending on the particular discretization
scheme, they may or may not be present as we will see below. Since the
eigenvalues of the corresponding continuous system are given by $\pm \i
\lambda_k$, where $\lambda_k$ is the k-th zero of the Bessel function
$j_1$, all the eigenvalues with nonvanishing real part are spurious,
i.e., they do not have an analogue in the continuous problem. Thus,
the corresponding eigenvectors are parasitic modes because they do not
approximate a regular mode.

When combined with a time integrator this spectrum and the stability
region of the integrator decide about the stability of the overall
scheme. The stability regions of some ODE solvers are sketched in
appendices~\ref{sec:stabreg} and~\ref{sec:icn}. Some of these include
an interval on the imaginary axis while others include only the
origin. All the stability regions extend further into the left half
plane than into the right. This implies that if a spectrum contains
points in the right plane (i.e., modes with positive real part) then
these modes tend to either unnecessarily diminish the time step
because they have to be scaled into the stability region (if the
stability region of ODE solver extends into the right half plane int
he first place) or else they create instabilities.

Another possibility to overcome instabilities is the addition of
numerical dissipation. This has the effect that the spectrum of
$\mathbf{A}$ is shifted towards the left. However, the presence of the
handles and of the outliers implies that the amount of dissipation
necessary for stabilising the scheme is rather high compared to cases
where the spectrum is located entirely on the imaginary axis. This
means that the evolution cannot be followed accurately because
ultimately the waves will be damped out. Thus, the appearance of
handles and outliers seems to leave us with the choice between the
Scylla of instability and Charybdis of inaccuracy. Therefore, we take
the appearance of these modes as a sign for an inadequate scheme.

The discretizations which we have looked at can be divided into three
groups. Discretizations of the system~\eqref{eq:system2nd} obtained
from the cartoon method of computing the transversal derivatives, 2nd
order discretizations of the system~\eqref{eq:adapted} and three
somewhat non-standard discretizations of that system. In the list
given below we only show the formula for the inner points of the
grid. The values at the boundary points are obtained by using the
appropriate symmetry conditions. Furthermore, we omit the equation for
$\dot v_i$ in those cases where its discretization is the same as for
the derivative term in the $\dot u_i$ equation.
\begin{enumerate}
\item Using the cartoon approximation we discretize the
  system~\eqref{eq:system2nd} to second order in $h$. The derivatives
  are taken to be centered in all the three different schemes, while
  we use different interpolations for the term $\frac{v_*}{x_*}$.
  \begin{enumerate}
    \renewcommand{\labelenumi}{(\alph{enumi})}
  \item Two point interpolation (see Fig.~\ref{fig:interp23}):
    \[
    \dot u_i = \frac{v_{i+1} - v_{i-1}}{2h} + \frac{(x_{i+1} - x_*) v_i
      + (x_* - x_i) v_{i+1}}{hx_*}.
    \]
  \item Three point interpolation (see Fig.~\ref{fig:interp23}):
    \[
    \dot u_i = \frac{v_{i+1} - v_{i-1}}{2h} + \sum_{j=-1}^{1}
    p_{i+j}(x_*) v_{i+j},
    \]
    where $p_{i-1}(x) = (x-x_i)(x-x_{i+1})/2h^2$, $p_i(x) =
    -(x-x_{i+1})(x-x_{i-1})/h^2$ and $p_{i+1}(x) = (x-x_i)(x-x_{i-1})/2h^2$.
    \begin{figure}[h]
      \begin{center}
        \includegraphics[width=7cm]{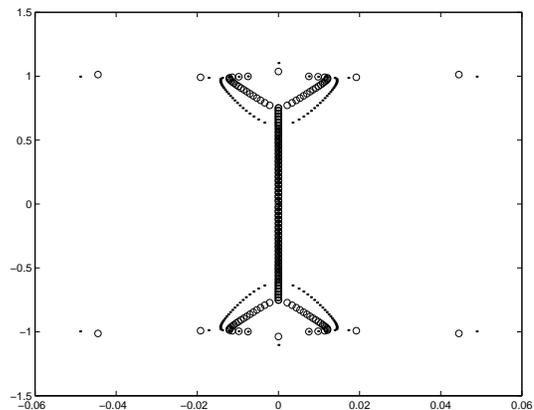}
        \caption{\label{fig:interp23} Two (dots) and three (circles)
          point interpolation of the cartoon method, schemes $1a$, $1b$}
      \end{center}
    \end{figure}
  \item Centered two point interpolation (see Fig.~\ref{fig:interp2c}):
    \[
    \dot u_i = \frac{v_{i+1} - v_{i-1}}{2h} + \frac{ (x_{i+1} - x_*) v_{i-1}
      + (x_*-x_{i-1}) v_{i+1}}{2hx_*}.
    \]
    \begin{figure}[h]
      \begin{center}
        \includegraphics[width=7cm]{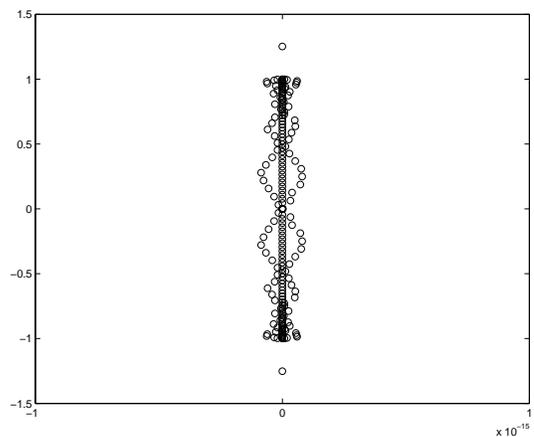}
        \caption{\label{fig:interp2c} Centered two point interpolation
          for the cartoon method, scheme $1c$}
      \end{center}
    \end{figure}
  \end{enumerate}
  It is obvious that the interpolation in schemes $1a$ and $1b$ is not
  useful because of the appearance of the handles and the
  outliers. This suggests that for higher order Lagrange interpolation
  one has to expect the same kind of behaviour. In fact, in our
  studies related to~\cite{frauendienerhein02:_numer} we find that
  e.g. 5th order Lagrange interpolation is as unstable as
  interpolation using Chebyshev polynomials. However, the fact that
  scheme $1c$ yields a purely imaginary spectrum (up to round-off
  error) suggests that the use of the centered interpolation is
  advantageous. In view of the fact that the cartoon method provides
  an approximation of the $1/x$ term in~\eqref{eq:adapted} we now look
  at this system directly.

\item Three second order discretizations of the system~\eqref{eq:adapted}.
  \begin{enumerate}
    \renewcommand{\labelenumi}{(\alph{enumi})}
  \item The scheme discussed already above (see Fig.~\ref{fig:straight})
    \[
    \dot u_i = \frac{v_{i+1} - v_{i-1}}{2h} + \frac{v_i}{ih}.
    \]
  \item A centered approximation for the numerator of the second term
    (see Fig.~\ref{fig:traditional2})
    \[
    \dot u_i = \frac{v_{i+1} - v_{i-1}}{2h} + \frac{v_{i+1} + v_{i-1}}{2ih}.
    \]
    \begin{figure}
      \begin{center}
        \includegraphics[width=7cm]{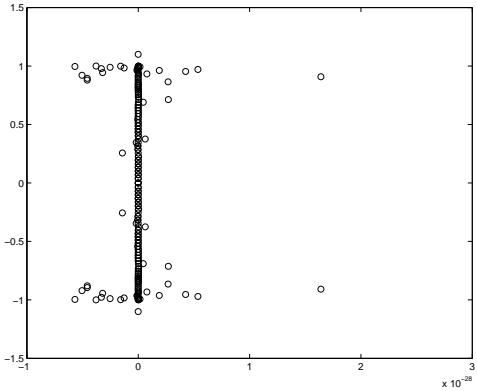}
        \caption{\label{fig:traditional2} Scheme $2b$}
      \end{center}
    \end{figure}
  \item A staggered grid discretization based on the variables
    $u_i = u(t,ih) $, $v_{i+1/2} = v(t,ih+h/2)$ (see Fig.~\ref{fig:staggered})
    \[
    \dot u_i = \frac{v_{i+1/2}-v_{i-1/2}}{h} + \frac{v_{i+1/2} +
      v_{i-1/2}}{2ih},\quad 
    \dot v_{i+1/2} = \frac{u_{i+1}-u_i}{h}.
    \]
    \begin{figure}[ht]
      \begin{center}
        \includegraphics[width=7cm]{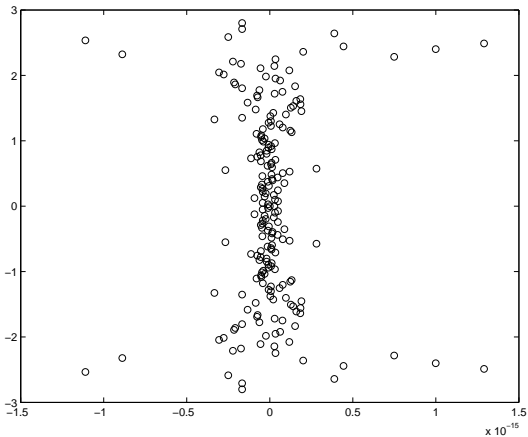}
        \caption{\label{fig:staggered}Scheme $2c$}
      \end{center}
    \end{figure}
  \end{enumerate}
  Again, we find that discretizing in a centered and compatible way
  yields a purely imaginary spectrum (again up to round-off
  error). The staggering in scheme $2c$ is suggested by the structure
  of the equations which is ultimately related to the meaning of the
  variables. While $u$ is a time derivative located at the grid points
  $v$ is a spatial derivative which should be located between the grid
  points.

\item The final three discretization schemes of~\eqref{eq:adapted} are
  somewhat different from the ones above:
  \begin{enumerate}
    \renewcommand{\labelenumi}{(\alph{enumi})}
  \item  This scheme takes the idea of centering to fourth order and again
    yields a purely imaginary spectrum (see Fig.~\ref{fig:fourth})
    \[
    \dot u_i = \frac{v_{i-2} -8 v_{i-1} + 8 v_{i+1} - v_{i+2}}{12h} +
    \frac{-v_{i-2} + 4 v_{i-1} + 4 v_{i+1} - v_{i+2}}{6ih}.
    \]
\begin{figure}[h]
      \begin{center}
        \includegraphics[width=7cm]{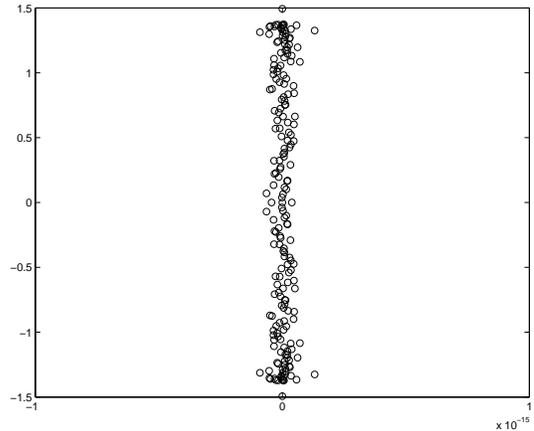}
        \caption{\label{fig:fourth}A fourth order centered scheme, $3a$}
      \end{center}
    \end{figure}
  \item The next scheme makes use of further information coming from
    the theory of group actions. While the radial coordinate $r$ is
    \emph{not} a smooth coordinate on the orbit space its square is
    smooth. Hence, it seems natural to express the variables as
    functions of the coordinate $x=r^2$. Then the minor redefinition $\tilde u(x)
    = 2u(r)$ and $\tilde v(x) = rv(r)$ of the variables yields the (regular)
    system
    \[
    \del_t {\tilde u} = \del_{x} \tilde v, \qquad 
    \del_t {\tilde v} = x \del_{x} \tilde u,
    \]
    which is discretized in the standard way (see Fig.~\ref{fig:r2standard})
    \[
    \dot {\tilde{u}}_i = \frac{\tilde v_{i+1} - \tilde v_{i-1}}{2h},
    \qquad
    \dot {\tilde {v}}_{i} = \frac{i}{2} (\tilde u_{i+1} - \tilde u_{i-1}).
    \]
    \begin{figure}[h]
      \begin{center}
        \includegraphics[width=7cm]{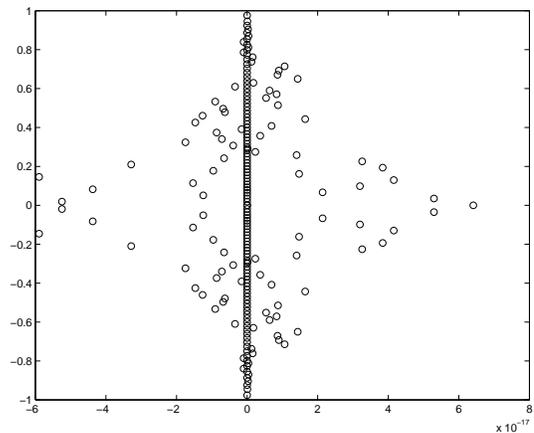}
        \caption{\label{fig:r2standard}Standard discretization for
          redefined system}
      \end{center}
    \end{figure}
  \item The same system discretized using a staggered grid (see
    Fig.~\ref{fig:r2staggered}) 
    \[
    \dot {\tilde{u}}_i = \frac{\tilde v_{i+1/2} - \tilde v_{i-1/2}}{h},
    \qquad
    \dot {\tilde {v}}_{i+1/2} = (i+1/2) (\tilde u_i - \tilde u_{i-1}).
    \]
    \begin{figure}[h]
      \begin{center}
        \includegraphics[width=7cm]{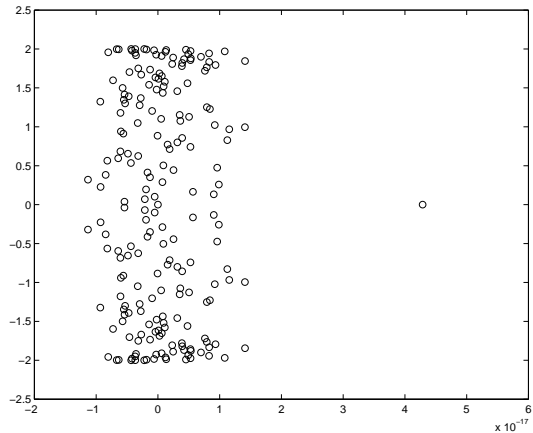}
        \caption{\label{fig:r2staggered}Staggered discretization for
          redefined system}
      \end{center}
    \end{figure}
  \end{enumerate}
\end{enumerate}

\section{Conclusion}
\label{sec:concl}

We discussed in this work several schemes for treating axisymmetric
systems which are difficult to handle numerically due to the degenarcy
of adapted coordinates and hence the formal singularity of the ensuing
equations. We applied various schemes to the axisymmetric wave
equation written as a symmetric hyperbolic system for the derivatives
and determined the spectrum of the ensuing matrices. As a consequence
of the shape of the stability regions of common ODE solvers we take
the appearance of handles and outliers described in
section~\ref{sec:numerics} as an indication for a `bad' discretization
scheme. These eigenvalues correspond to grid modes which, when evolved
with a standard ODE solver, will lead to instabilities. In order to
avoid them one needs to be careful in discretizing the equations
because the derivative term and the $1/x$ term must be treated in a
compatible way. This is due to their common geometric origin. In the
general case, where tensors of higher rank are involved different
terms may have to be combined. Exactly which terms these are is
indicated by the behaviour of the tensor components under the symmetry
transformation.

One particular method we discussed is the cartoon method. It turns out
that this method is not really different from a direct implementation
of the axisymmetry in terms of the adapted coordinates. Here the
problem is to make the interpolation compatible with the
discretization in the $y=0$ half plane so that parasitic modes are
excluded. However, the advantage in using the cartoon method is that
it admits a quick and easy extension of a 2D axisymmetric code to a
full 3D code because the underlying grid structure and the equations
need not be changed.

Clearly, this investigation cannot deliver the ultimate solution to
the problem. It should be regarded as providing a guide to eliminate
`bad' discretizations and suggesting possible `good' ones. For instance, we
have implemented the schemes $1c$ and $3a$ in our axisymmetric
code~\cite{frauendienerhein02:_numer} for solving the conformal field
equations. While these schemes perform well for the axisymmetric wave
equation they still tend to produce slowly growing instabilities in
the case of the quasilinear conformal field equations. This behaviour
might be related to the fact that in most cases the real parts of the
eigenvalues vanish only up to round-off error and that these small
real parts start to grow due to non-linear interactions. It would be
useful to have a scheme in which the spectrum is \emph{exactly} on the
imaginary axis and whose only non-vanishing eigenvalues are the
physical ones. However, such a scheme is probably impossible to find.

\begin{acknowledgments}
This work was supported in part by NATO Collaborative Linkage
grant PST.CLG.978726.   
\end{acknowledgments}

\begin{appendix}
  
\section{Stability regions}
\label{sec:stabreg}

We illustrate the procedure for determining the stability region of a
particular time evolution scheme for an explicit Runge-Kutta
scheme. Let $\dot u = f(t,u)$ be the (system of) ODE(s) which we
intend to solve. An explicit Runge-Kutta scheme with $s$ `stages' to
compute an approximation $u_1$ of $u(t_0+h)$ is formulated as
follows:
\begin{align*}
  k_1 &= f(t_0,u_0),\\
  k_2 &= f(t_0 + c_2 h, u_0 + h a_{21} k_1),\\
  k_3 &= f(t_0 + c_3 h, u_0 + h a_{31} k_1 + h a_{32} k_2),\\
&\cdots\\
  k_s &= f(t_0 + c_s h, u_0 + h a_{s1} k_1 + h a_{s2} k_2 + \cdots + h
  a_{ss-1} k_{s-1}),\\
  u_1 &= u_0 + h (b_1 k_1 + \cdots + b_s k_s).
\end{align*}
The particular scheme is characterized entirely by the coefficients
$a_{ik}$, $b_i$ and $c_k$ which are conveniently presented as a
tableau
\[
 \setlength{\arraycolsep}{10pt}
\renewcommand{\arraystretch}{0.3}
\begin{array}{c|ccccc}
0  &        &        &        &         &\\
c_2& a_{21} &        &        &         &\\
c_3& a_{31} & a_{32} &        &         &\\
\vdots&\vdots&\vdots&\ddots&&\\
c_s& a_{s1} & a_{s2} & \cdots & a_{ss-1}&\\
\hline
   & b_1    & b_2    & \cdots & b_{s-1} & b_s
\end{array}
\]
The method is said to be of $p$-th order if for sufficiently smooth
$f$ the condition
\[
||u_1 - u(t_0+h)|| \le C h^{p+1}
\]
holds.

To get to the stability region one applies the scheme to the specific
model equation $\dot u = \lambda u$ where $\lambda\in\CC$ is
arbitrary. Going through the procedure above it is clear that the
final result is of the form
\[
u_1 = P(\lambda h) u_0
\]
where $P(z)$ is a polynomial of degree $s$, the so called stability
function of the method. Since $||u_1|| = |P(\lambda h)|\; ||u_0||$ we
find that the method will produce bounded approximations if and only
if $|P(\lambda h)| \le 1$. This motivates the definition of the
stability region as the set
\[
\Omega = \{ z\in \CC \bigm\vert \;|P(z)| \le 1\}.
\]
In case of an implicit scheme the stability function, obtained in the
same way by applying the scheme to the model equation, is not a
polynomial but a rational function. In a similar way, the stability
regions of linear multistep schemes are obtained. Here, the simplest
one is the leapfrog scheme which updates $u_{n+1}$ from the values on
the previous time levels $u_n$ and $u_{n-1}$ as $u_{n+1}=u_{n-1} +
2 h f(t_0+nh,u_n)$. It is easy to see that the stability region for
the leapfrog scheme is the part of the imaginary axis between $-\i$ and
$\i$.

\begin{figure}
  \begin{center}
    \includegraphics[width=4cm]{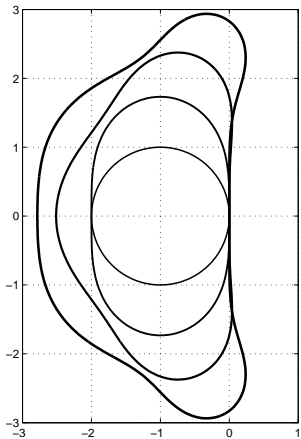}
    \caption{\label{fig:stabreg1} Stability regions for explicit
      Runge-Kutta schemes with $p=s$}
  \end{center}
\end{figure}

\begin{figure}[ht]
  \includegraphics[width=4cm]{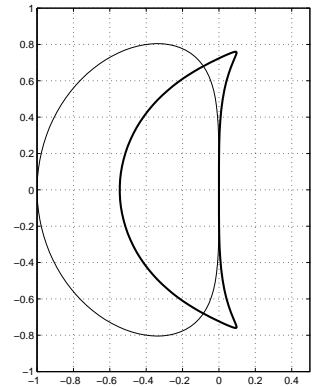}
  \caption{\label{fig:ab23}Stability regions for second (thin line)
    and third (thick line) order Adams-Bashforth methods} 
\end{figure}

The stability regions for explicit Runge-Kutta schemes of order $p=s$
for $s=1,2,3,4$ are shown in Fig.~\ref{fig:stabreg1}.
The stability regions for the second and third order Adams-Bashforth
methods are shown in Fig.~\ref{fig:ab23}
as an example for stability regions of multistep methods. For the
stability regions of other multistep schemes we refer
to~\cite{fornberg96,gustafssonkreiss95:_time}. Qualitatively, these
regions extend into the left half plane of the complex plane including
possibly some part of the imaginary axis. For higher order methods the
stability region may even extend into the right half plane.

\section{The ICN time evolution scheme}
\label{sec:icn}

\begin{figure*}
   \mbox{ 
     \includegraphics[width=3cm]{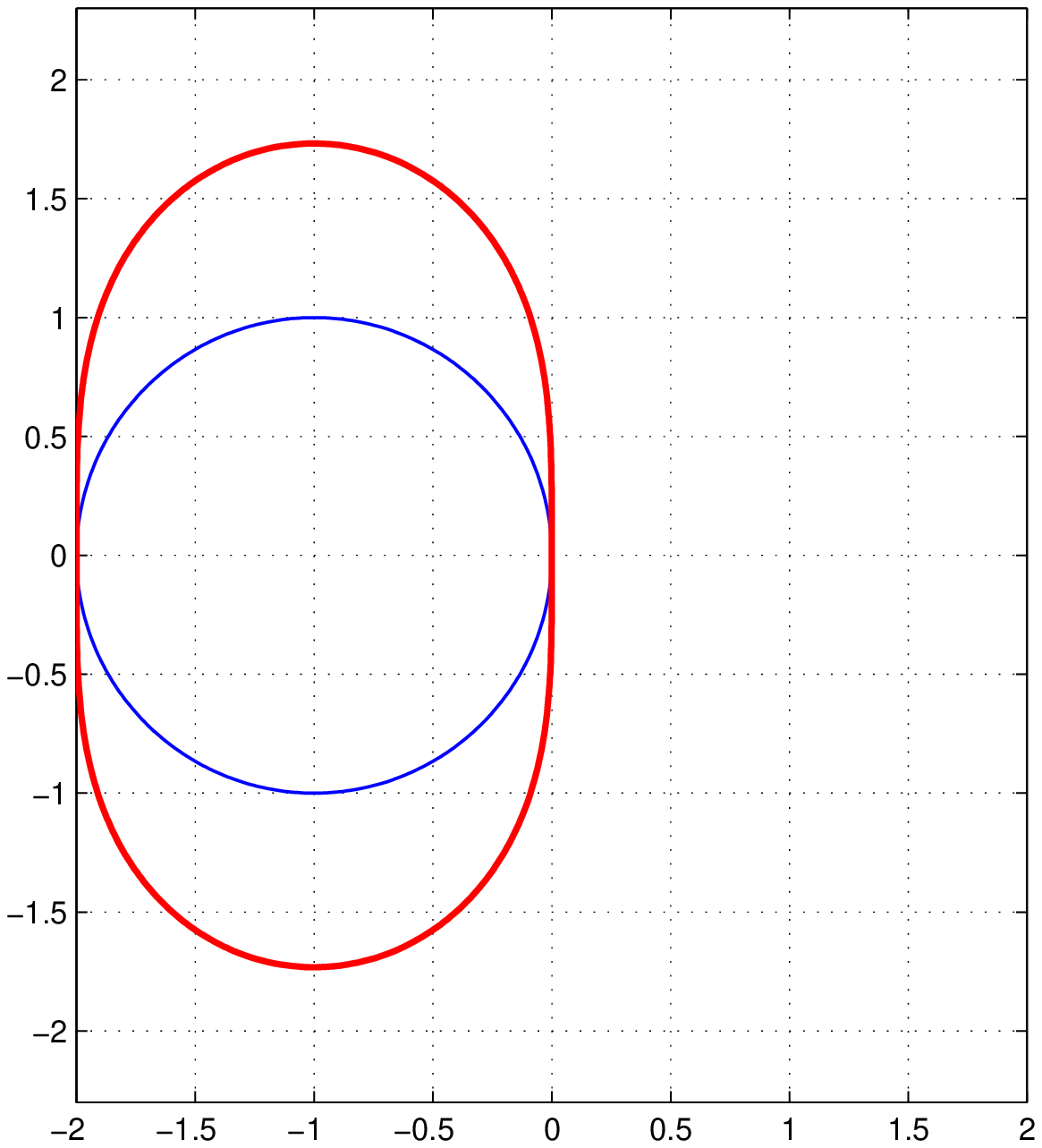}
     \hfill
     \includegraphics[width=3cm]{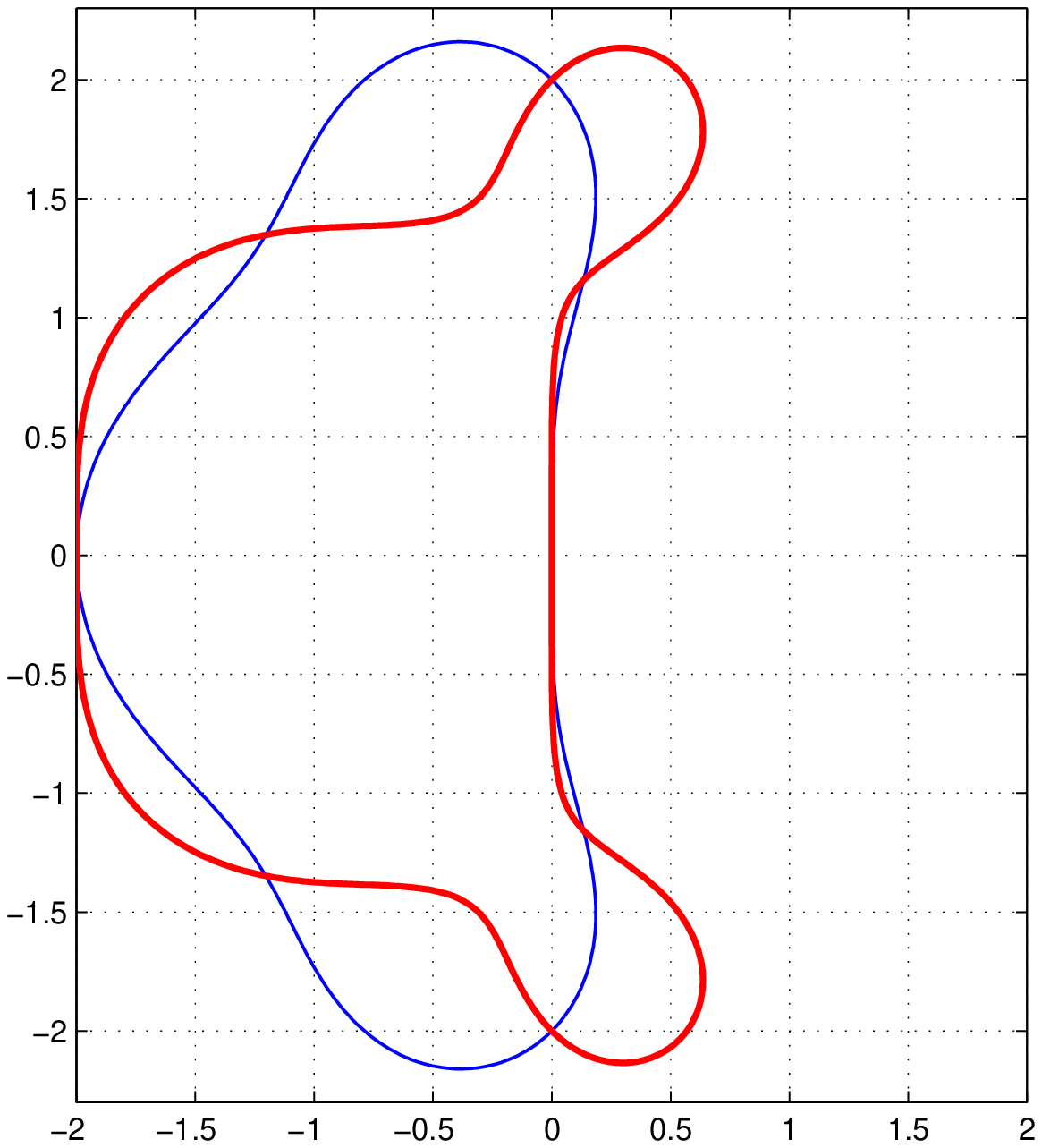}
     \hfill
     \includegraphics[width=3cm]{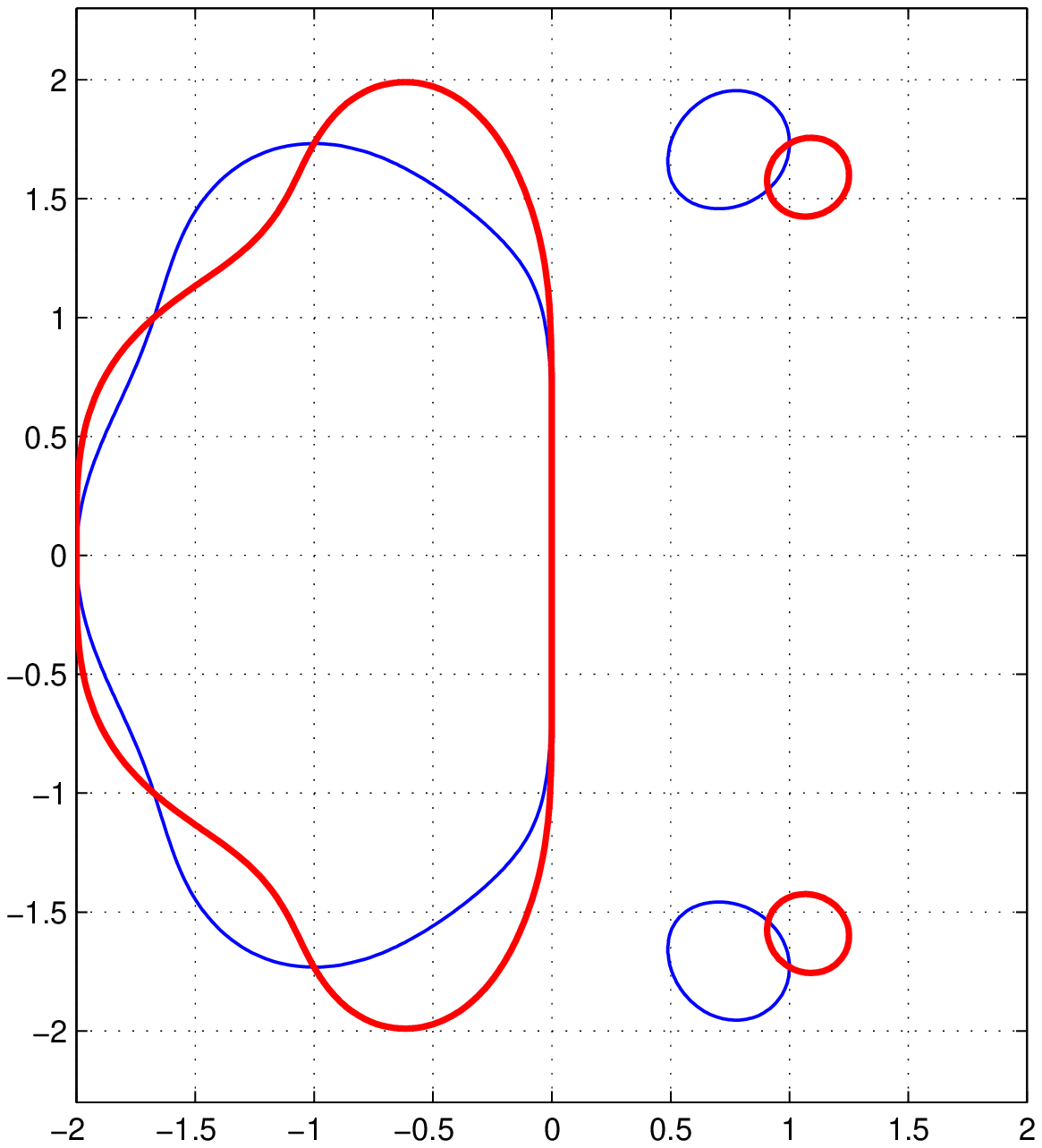}
     \hfill
     \includegraphics[width=3cm]{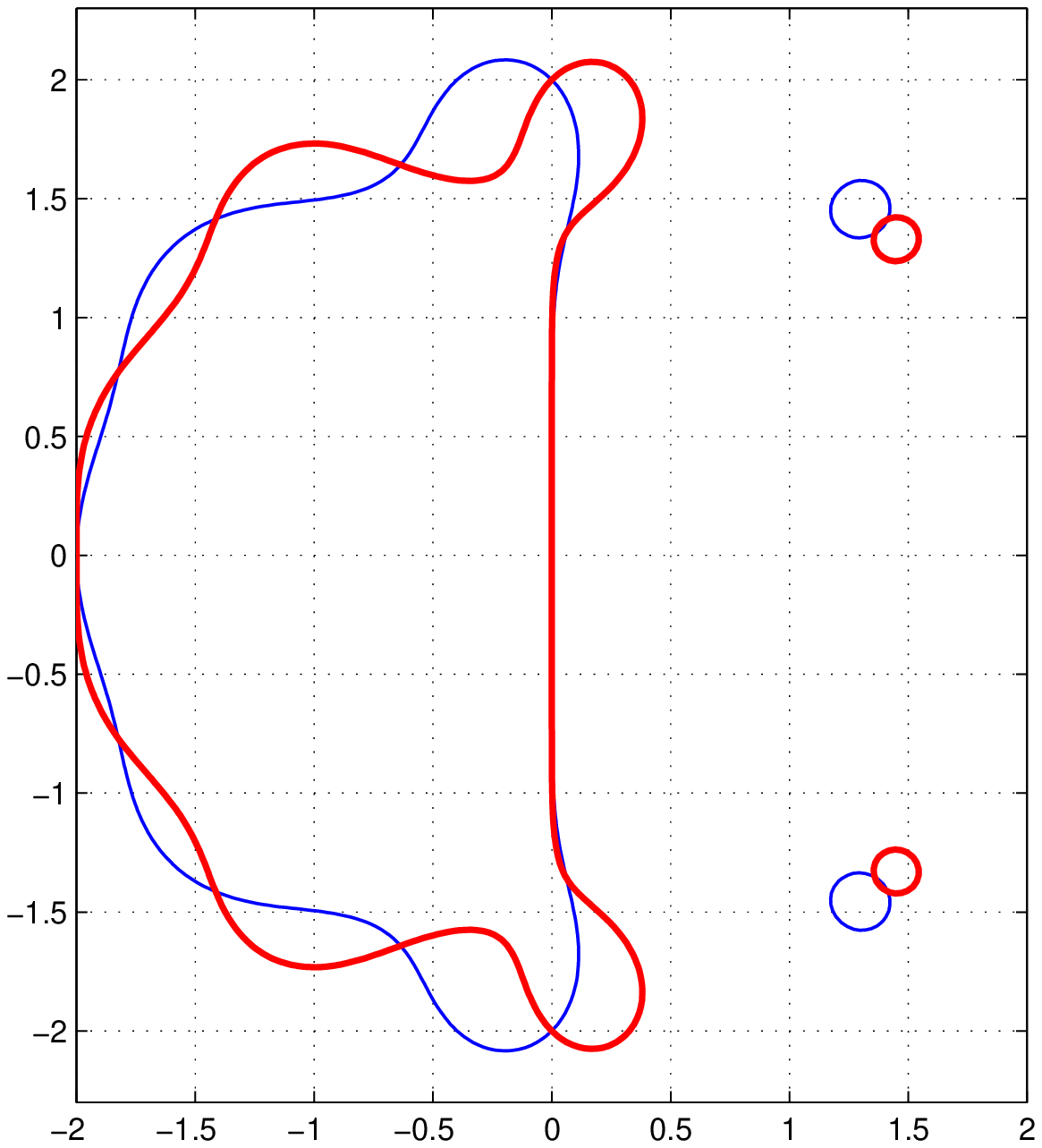}
     }
    \caption{\label{fig:icn}The stability regions for ICN(s) for
      $s=1,2$, $s=3,4$, $s=5,6$ and $s=7,8$}
\end{figure*}

The ICN scheme was described
in~\cite{teukolsky99:_crank_nichol_method_numer_relat} in application
to a one-dimensional advection equation. Viewing the ICN scheme as a
time stepper to solve systems of ODEs $\dot u = f(t,u)$ one can deduce from
this description the following algorithm to compute $u_1 \approx u(t_0+h)$
\begin{align*}
  k_1 &= f(t_0,u_0),\\
  k_2 &= f(t_0 + \2 h, u_0 + \2 h k_1),\\
  k_3 &= f(t_0 + \2 h, u_0 + \2 h k_2),\\
&\cdots\\
  k_s &= f(t_0 + \2 h, u_0 + \2 h k_{s-1}),\\
  u_1 &= u_0 + h k_s.
\end{align*}
Cleary, this is an explicit Runge-Kutta scheme with the tableau
\[
\setlength{\arraycolsep}{10pt}
\renewcommand{\arraystretch}{0.3}
\begin{array}{c|ccccc}
0  &    &    &        &         &\\
\2 & \2 &    &        &         &\\
\2 &  0 & \2 &        &         &\\
\vdots&\vdots&\vdots&\ddots&&\\
\2 &  0& 0 & \cdots & \2&\\[5pt]
\hline
   &  0  &  0   & \cdots & 0 & 1.
\end{array}
\]
Note, that $k$ iterations correspond to $s=k+1$ stages of the
corresponding Runge-Kutta scheme. Application of this scheme to the
model equation $\dot u = \lambda u$ yields the stability function
\[
P_s(z) = 1 + z \sum_{k=0}^{s-1} \left(\frac{z}{2}\right)^k = 1 + z +
\frac{z^2}{2} + \frac{z^3}{4} \sum_{k=0}^{s-3} \left(\frac{z}{2}\right)^k.
\]
The fact that the first three terms agree with the Taylor expansion of
$\exp(z)$, the stability function of the exact time evolution, shows
that this scheme is second order
accurate~\cite{hairernorsett00:_solvin_ordin_differ_equat}. The
stability regions are obtained as the level set $\{P_s(z) P_s(\bar
z)=1\}$. The regions for some values of $s$ are shown in
Fig.~\ref{fig:icn}.
We see that the imaginary axis intersects the stability regions for
$s=3,4,7,8,\ldots$ in a finite interval while for the other values
they have only the origin in common with the imaginary axis. To
explain this somewhat unexpected result we look at the behaviour of
the modulus of the stability function on the imaginary axis in a
neighbourhood of the origin. Writing
\[
P_s(z) = 1 + z \sum_{k=0}^{s-1} \left(\frac{z}{2}\right)^k =
1+z\frac{1-\left(\frac{z}{2}\right)^{s}}{1-\frac{z}{2}}
\]
and inserting $z=\i y$ into $Q(z) = P_s(z) P_s(\bar z)$ we find
\begin{widetext}
\[
Q(\i y) = 1 + 2 \frac{-(\i y/2)^{s+1} (1+(-1)^{s+1}) +
  (\i y/2)^{s+2}(1+(-1)^{s}) + 2 (y/2)^{2s+2} }{1+(y/2)^2}.
\]
\end{widetext}
This shows that for all $s > 1$
\[
Q(\i y) = 
\begin{cases}
  1 - \left(\frac{\displaystyle \i}{\displaystyle 2}\right)^s\, y^{s+2}
  + \cdots & s\text{ even}\\
  1 + \left(\frac{\displaystyle \i}{\displaystyle 2}\right)^{s+1}\, y^{s+1}
  + \cdots & s\text{ odd}.
\end{cases}
\]
The behaviour of the stability regions is determined by the
coefficient of the lowest order term. These lead to the sequence $1$,
$1/4$, $-1/4$, $-1/16$, $1/16$, $\dots$ for $s=1,2,\ldots$. The
negative coefficients imply a decrease of $Q(iy)$ along the
imaginary axis close to $y=0$ and hence they correspond to the cases
where the stability region contains a part of the imaginary axis. The
other cases, when the coefficient of the dominating term is positive,
are those when the stability region just touches the imaginary axis at
the origin. In these cases, systems with eigenvalues on or to the
right of the imaginary axis such as in particular the advection
equation and other hyperbolic equations cannot be stably evolved with
ICN. Thus, we come to the same conclusion as
in~\cite{teukolsky99:_crank_nichol_method_numer_relat}, namely that
ICN($k$) for $k=2,3,6,7,\ldots$ is unstable for hyperbolic
equations. Furthermore, since the method is always second order
accurate regardless of how many iterations are performed, there is no
use in performing more than two. In fact, in most practical cases it
is probably more efficient to use a higher order Runge-Kutta method
instead.

Another interesting aspect of the ICN scheme is its relationship with
the Crank-Nicholson method. This is a second order accurate, implicit
time stepping scheme whose stability function is given by
\[
P(z) = \frac{1+z/2}{1-z/2}
\]
and whose stability region is the entire left plane. Taking the limit
of $P_s(z)$ for $s\to \infty$ we find that in fact $P_s(z) \to
P(z)$. However, absolute convergence holds only in the circle $|z| <
2$. This has the following implication: iterating the ICN scheme until
convergence to produce approximations to the exact Crank-Nicholson
method works only within that circle. Thus, this scheme (which might
be termed `the full ICN scheme') also has a stability barrier and
hence a limitation of the time step in contrast to the exact CN scheme
which is unconditionally stable. Thus, we conclude that the ICN scheme
is not an efficient alternative to higher order Runge-Kutta or
implicit schemes like CN.
\end{appendix}

\end{document}